\documentclass{webofc}
\usepackage[varg]{txfonts}   
\def\lsim{\mathrel{\rlap{\lower4pt\hbox{\hskip1pt$\sim$}}
    \raise1pt\hbox{$<$}}}
\def\gsim{\mathrel{\rlap{\lower4pt\hbox{\hskip1pt$\sim$}}
    \raise1pt\hbox{$>$}}}
\def\sqr#1#2{{\vcenter{\vbox{\hrule height.#2pt
         \hbox{\vrule width.#2pt height#1pt \kern#1pt
         \vrule width.#2pt}
         \hrule height.#2pt}}}}

\def\laq{\raise 0.4 ex \hbox{$<$}\kern -0.8 em\lower 0.62 ex\hbox{$\sim$}}
\def\gaq{\raise 0.4 ex \hbox{$>$}\kern -0.7 em\lower 0.62 ex\hbox{$\sim$}}

\def\be{\begin{equation}}
\def\ee{\end{equation}}
\def\beqa{\begin{eqnarray}}
\def\eeqa{\end{eqnarray}}

\def\dalemb#1#2{{\vbox{\hrule height.#2pt
        \hbox{\vrule width.#2pt height#1pt \kern#1pt \vrule width.#2pt}
        \hrule height.#2pt}}}

\def\dalemb#1#2{{\vbox{\hrule height.#2pt
        \hbox{\vrule width.#2pt height#1pt \kern#1pt \vrule width.#2pt}
        \hrule height.#2pt}}}

\def\gtorder{\mathrel{\raise.3ex\hbox{$>$}\mkern-14mu
             \lower0.6ex\hbox{$\sim$}}}
\def\ltorder{\mathrel{\raise.3ex\hbox{$<$}\mkern-14mu
             \lower0.6ex\hbox{$\sim$}}}

\woctitle{The Time Machine Factory}
\begin{document}
\title{Zen and the Art of Space-Time Manufacturing}
%
%

\author{Orfeu Bertolami\inst{1,2}\fnsep\thanks{\email{orfeu.bertolami@fc.up.pt}} 
}

\institute{Departamento de F\'\i sica e Astronomia, 
Faculdade de Ci\^encias, Universidade do Porto
\and
Instituto de Plasmas e Fus\~ao Nuclear, Instituto Superior T\'ecnico, Universidade T\'ecnica de Lisboa           
 }

\abstract{%
We present a general discussion about the so-called emergent properties and discuss whether space-time and gravity can be regarded as emergent features of underlying more fundamental structures. Finally, we discuss some ideas about the multiverse, and speculate on how our  universe might arise from the multiverse.}
\maketitle
\section{Introduction}
\label{intro}
The reductionist approach has shown to be extremely useful to unravel the fundamental building blocks of matter and has lead to fecund discoveries in physics, chemistry and biology. It is a method that focus not only on the most elementary building blocks of matter, but also on the interactions that bind this basic pieces at a given scale of energy. The reductionist approach aims fundamentally to explain the complexity of Nature by the motion and interaction of elementary building blocks. It is an approach does not preclude the existence of the so-called emergent properties or phenomena, even though it is clearly based on the assumption that these phenomena can be {\it entirely} explained by the underlying interactions between the elementary building blocks. Some claim that the reductionist perspective can be more than a theoretical and a methodological procedure and can acquire an ontological dimension, if reality itself, can be decomposed in a minimum set of entities (see e.g. Ref. \cite{Jones} for a detailed discussion).  

The reductionist perspective can be seen from different points of view, such as recently suggested, for instance, by Wolfram, according to whom complex phenomena and patterns can emerge from simple computational rules, the cellular automata \cite{Wolfram}, a perspective that has been criticized by many accounts (see e.g. Ref. 
\cite{Weinberg1}).

In direct opposition to reductionism one encounters what has been generically referred to as {\it emergence}. Emergence is the way complex systems and patterns arise out of a multiplicity of simple interactions and manifests itself in physics, biology, philosophy, economics, architecture, etc. One usually distinguishes two forms of emergence: Weak emergence, a model or a linguistic way to refer to new properties arising as a result of the interactions at an elemental level; Strong emergence, when the new emergent properties are irreducible to their constituent parts, when the whole is greater than the sum of its parts  (see e.g. the discussion of Ref. \cite{Laughlin}).

The difference between these contrasting approaches is not relevant for us, and we shall focus instead, on the way these ideas might loosely help us in bringing insights about the nature of space, time and gravity, whether these are fundamental building blocks or, actually, emergent properties of more fundamental underlying substances. 
Naturally, these are not easy questions and, in fact, there is very little in sight to suggest that we must go beyond the current perspective that space-time, although plastic and mutable (see e.g. Ref. \cite{Bertolami06} for a broad account), is the basic setting of all phenomena and is not derivable from any more elementary entities. The same could be said about gravity, one of the four fundamental interactions of Nature identified so far. But, of course, the apparent elemental nature of space-time and gravity might, in fact, reflect our ignorance about a putative ultimate theory of everything (see e.g. Refs. \cite{Weinberg2,Bertolami2010} for general discussions). 

Indeed, through the development of statistical mechanics and quantum theory it has been possible to understand, for instance, that temperature is a composite quantity, actually connected with the kinetic energy of the atoms and molecules. The same can be said about irreversibility. Thus, temperature and irreversibility are emergent properties, features that can be understood from the underlying behaviour of matter at a more fundamental level. And in fact, the reductionist approach allows one to establish, through the formalism of statistical and quantum physics, that many quantities are derivable from more fundamental quantities and/or constants. Hence, the issues we aim to discuss here are whether space-time and gravity are emergent properties? That is, is space-time derivable from more fundamental entities? Is gravity an emergent property? In what follows we shall review some of the currents of thought about these questions.

\section{Before space-time}
\label{space-time}

\subsection{Pregeometry}

Pregeometry is reminiscent from the ideas of Wheeler about ``it from bit" \cite{Wheeler}, and more concretely due to the insight of Sakharov about the metric elasticity of space \cite{Sakharov}. In this approach it is assumed that the relationship between matter and curvature as established by general relativity (GR) is due to some underlying set of heavy fields and some features of space itself. In the words of Sakharov: 

\noindent
"The presence of the GR action leads to a metrical elasticity of space, i.e., to generalized forces which oppose the curving of space. Here we consider the hypothesis which identifies the GR action with the change in the action of quantum fluctuations of the vacuum if space is curved."    

Thus, if one assumes that heavy fields are the underlying support of GR, then pregeometry or induced gravity, as it was often called, is a  programme to compute the induced Newton's constant, $G_{ind}$, and the induced cosmological constant, $\Lambda_{ind}$, from a renormalizable theory of preogeometric fields (see e.g. \cite{Akama} and Ref. \cite{Adler} for a review). And indeed, it can be shown that GR plus higher-order curvature terms can be obtained from the functional integration of the pregeometric heavy fields; however, quite general field theory arguments show that $G_{ind}$ and $\Lambda_{ind}$ cannot be unambiguously computed (see e.g. Ref. \cite{David}).

\subsection{Loop Quantum Gravity and Ho\v{r}ava-Lifshitz Gravity}

Of course, the most fundamental nature of space-time requires the understanding of physics at the Planck scale and the how quantum effects can be made consistent with a covariant picture for the space-time geometry. For instance, in the loop-quantum gravity proposal, the space-time picture that emerges from the formalism is a ``weaving'' of Planck size fundamental loops (see e.g. ref. \cite{Abhay} for a status report). 

We shall discuss some ideas arising from string theory in the next subsection, but before that we briefly consider the so-called Ho\v{r}ava-Lifshitz (HL) gravity, a quite original proposal for a ultraviolet (UV) completion of GR \cite{Horava:2009uw}. In this approach space and time are treated in a different way at the UV regime and this renders the resulting gravity model power-countable renormalizable. GR is supposed to be recovered at the infra-red (IR) fixed point. Thus, in order to obtain a renormalizable gravity theory one abandons Lorentz symmetry at high-energies \cite{Horava:2009uw,Visser:2009fg}. Even though the idea that Lorentz symmetry is a low-energy symmetry has been previously considered in field theory \cite{Chadha:1982qq} and string theory \cite{Kostelecky}, the novelty of the HL proposal is that the breaking of Lorentz symmetry occurs, likewise in some condensed matter models (cf. Ref. \cite{Horava:2009uw} and references therein),  through an anisotropic scaling of space and time: $\vec{r}\rightarrow b\vec{r}$ and $t \rightarrow b^{z} t$, $b$ being a scale parameter and $z \not= 1$. Lorentz symmetry is presumably recovered at a IR fixed point, and $z$ flows to $z=1$ in this limit.

The anisotropy between space and time leads to the well known $3+1$ Arnowitt-Deser-Misner (ADM) splitting \cite{Arnowitt:1962hi}, originally devised to express GR in a Hamiltonian form. Following Ref. \cite{Horava:2009uw}, a foliation,  parametrized by a global time $t$, can be introduced through a weaker form of the global diffeomorphism, the so-called  {\it foliation-preserving diffeomorphism}. Despite its potentially interesting features, HL gravty is afflicted by various illnesses 
(see e.g. Ref. \cite{Horava:2009uw}). Nevertheless, although it does bring some relevant insight for high energy cosmology. Indeed, cosmological considerations have been extensively studied in HL gravity (for reviews, see Refs. \cite{Mukohyama:2010xz,Saridakis:2011pk}), some of which in the context of quantum cosmological models \cite{Bertolami-Zarro}.

\subsection{Emergent space-time from string theory}

In this subsection we discuss the striking possibility that space-time arises from string theory. We think that it is fair to state that the status of the subject does not allow for claiming that there is a conclusive connection, but arguments are, at least, quite compelling. Most of these are encountered in the persuasive paper of Nathan Seiberg \cite{Seiberg}.

The first point made by Seiberg concerns the ambiguous way space can be probed in classical string theory, given the extended nature of strings and hence, the impossibility of probing arbitrarily short distances. This becomes quite concrete through the so-called T-duality, a symmetry of the string mass spectrum, according to which the mass of strings states compactified on a circle of radius $R$, is exactly the same if on a circle of radius $\alpha'/R$, where $\alpha'$ is dimensional constant that characterizes the string, so that $\sqrt{\alpha'}$ is the typical string length scale, $l_s$, presumably close to the Planck length, $l_P =\sqrt{G\hbar/c^3} \simeq 10^{-35}~m$. 

A similar feature is found when considering orbifolds. An orbifold is a generalization of a manifold that allows the presence of points whose neighborhood is diffeomorphic to the quotient of $\bf{R}^n$ by a finite group. In physics, the notion of an orbifold usually describes an object that can be globally written as an orbit space $M/\Gamma$, where $M$ is a manifold and $\Gamma$ is a group of its isometries; hence, it follows that states on a circle of radius $R$ are the same as its orbifold with a typical linear length, $R$. 

Moreover, space is essentially ambiguous at the Planck length as in order to resolve distances, $r$, smaller the than the string length, $l_s$,  energy must be concentrated such that $E>1/r$, but this creates a black hole, unless $r	>l_P$. Of course, this gives rise to further ambiguities and so do changes on the string coupling due to s-duality, branes exchange, etc. In other words, these ambiguities indicate that higher energies do not imply in a better resolution, but rather, they turn the probe bigger.

The next point concerns the fact that general covariance is actually a derived symmetry. In order to understand that several facts can be evoked. From the start, it is clear that  general covariance is a redundancy in the physical description, akin to gauge symmetry, even though not a symmetry of the Hilbert space; however, duality symmetries that are so crucial in string theory indicate that gauge symmetries are not fundamental, and hence it suggests that general covariance is not fundamental either. But without general covariance,  the energy-momentum tensor cannot be constructed and space-time itself does not seem to be fundamental. This all means that one cannot talk about local observables and, perhaps, even local degrees of freedom. In fact, even locality is under question if no local observables strictly exist. 

A further ambiguity arises from the CFT/AdS conjecture \cite{Maldacena}, according to which a string theory in AdS backgrounds is dual to a local quantum field theory on the boundary, meaning that conformal field theory is holographic to the bulk string theory. So this gauge/gravity duality implies that locality is, in strict terms, lost. 

The argument put forward by Seiberg is completed once D-branes (D0 branes, actually) are considered. Indeed, D-brane dynamics is such that only the relative position between branes is meaningful. This implies that space does made sense only in the presence of branes. Thus, the fundamental conclusion is that space does not seem to be a fundamental entity and, if time is at the same footing than space, hence both space and time are emergent properties from the most basic entities of string theory.

\section{Time traveling} 

Let us now turn to the main subject of the Time Machine Factory conference. Thanks to relativity, whose experimental adequacy and accuracy is in agreement of all known data (see eg. Ref. \cite{BP2012}), space and time are seem as different aspects of the same entity (for a brief survey of possible alternatives to GR, see e.g. Ref. \cite{Bertolami2011}). However, the existence of several arrows of time, which are clearly correlated to each other, suggests that time has some attributes that are not shared by space (for discussions see e.g. Refs. \cite{Bertolami08,Bertolami-Lobo}) and one can loosely speak about a ``phenomenology" of time. Indeed, speculations about the nature of time are as old as the history of rational thinking, and in this respect, many notions of time can be mentioned:

\vspace{0.3cm}

\noindent
i) The notion that time is {\it linear} and that events do not repeat themselves as defended by Heraclitus, Aristotle, St. Augustin, Maimonides, the medieval rabbi Isaac Luria, and many others.

\vspace{0.3cm}

\noindent
ii) Ciclic time, a believe of many ancient civilizations, and which, in GR, correspond to space-time solutions of Einstein's field equations that contain closed timelike curves (CTCs). Philosophically, this notion has been discussed by Spencer Nietzsche and is associated to so-called  ``eternal return" of all things. This discussion is in physics associated with Poincare's recurrence theorem in the phase-space. 

\vspace{0.3cm}

\noindent
iii) Historical time, which has its roots in the thinking of the philosopher Henry Bergson, who defended that the time of living beings is not the mathematical time of the space-time formulation of relativity. Actually, this view does arise in discussions about time traveling, and is usually referred to as {\it predestination assumption}: 
\vspace{0.3cm}

\noindent 
A time traveller attempting to alter his/her past, intentionally or not, would only be fulfilling his/her role in creating a known history, but not changing it. That is, the time-traveller knowledge of its own history already includes future time travels.

\vspace{0.3cm}

\noindent
iv) The ``emancipated'' time, after the Ho\v{r}ava-Lifshitz proposal for gravity, in which, as we have seen, at high energies or at small scales, space and time are 
{\it different}.

\vspace{0.3cm}

\noindent
v) Finally, we mention a ``post-modern'' time proposal, the ``GPS'' time. This is a particularly interesting byproduct of the GPS system, according to which the mutual knowledge of four proper times allow for establishing the position of the clocks everywhere in the Universe \cite{Rovelli,Coll} (see also the contribution of Angelo Tartaglia in this volume).

\vspace{0.3cm}
After this characterization, let us now turn the discussion to CTCs. In the context of GR solutions a typology includes: 
\vspace{0.3cm}

\noindent
Solutions in which a piece of the light cone is missing due to rotation \cite{Godel}, or due to the presence of cosmic strings \cite{Gott}, whose physical variables often involve an unphysical equation of state \cite{DeserJackiw,Deser}.

\vspace{0.3cm}

\noindent
Traversable wormhole solutions (see e.g. Refs. \cite{Morris-Thorne-Yurtsever, Lobo}). 

\vspace{0.3cm}

\noindent
Warp drive solutions (see e.g. Refs. \cite{Alcubierre,Lobo-Visser}).

\vspace{0.3cm}

\noindent
The so-called Krasnikov tube-type solutions \cite{Krasnikov}.

\vspace{0.3cm}

Of course, the existence of CTCs give origin to several paradoxes, which represent a serious threat to causality. The most well known of them is the 
{\it killing of an ancestor paradox}, but one could also refer to the {\it creation of information paradox}: For instance, a time traveller from the future conveys the secret of time traveling to a researcher. Later, the researcher travels back in time and convey the secret to his/her younger person. So the information seems to have appeared from nowhere. Therefore, in order to avoid these paradoxes, one expects that CTCs are excluded, either by logical or by physical arguments, or can be accommodated without conflict with causality. We mention some of the suggested solutions. 

\vspace{0.3cm}

\noindent
Novikov's {\it self-consistent principle} which states that \cite{Novikov}:

\vspace{0.3cm}

\noindent
CTCs might exist, but they cannot entail any type of causality violation or time paradox, so that there is only one timeline, or that alternative timelines (such as in the many-worlds interpretation (MWI) of quantum mechanics (QM)) are not accessible.  

\vspace{0.3cm}

\noindent
Another criteria was proposed by Stephen Hawking, the {\it chronology protection conjecture} \cite{Hawking1992}: 

\vspace{0.3cm}

\noindent
``It seems that there is a Chronology Protection Agency which prevents the appearance of CTCS so to make the universe safe for historians.''

\vspace{0.3cm}

\noindent
What does QM and, in particular the MWI has to say about CTCs? The issue has been investigated in Ref. \cite{Deustch}, it is concluded that systems can travel from one time in one world to another time in another world; however, no system can travel to an earlier time in the same world.

\vspace{0.3cm}

\noindent
Finally, let us present a suggestion which has its roots in the time machine solution \cite{Bertolami-Ferreira} (see also the contribution of Ricardo Zambujal Ferreira in this volume) encountered in the context of f(R) theories with non-minimal coupling between curvature and matter \cite{Bertolami-Boehm-Harko-Lobo}, but that is actually found in many other solutions (see e.g. Ref. \cite{Costa}). Our argument is that CTCs do not actually exist since they require: in GR, conditions for matter that are unphysical (see e.g 
\cite{DeserJackiw,Deser}); and beyond GR, in regimes in which the alternative models of gravity are themselves, no longer valid. 

Let us close this discussion and point out that if CTCs are possible, then reversibility is reestablished; however, since irreversibility is a distinct feature of our world, then it seems logical to conclude that gravity itself must be an emergent property and CTCs are not possible. In what follows we review a recent proposal for the emergence of gravity.

\section{Emergent Gravity}

Recently, it has been proposed that Newton's inverse square law (ISL) and Einstein's field equations can be derived from thermodynamical considerations and from the Holographic Principle \cite{Verlinde}. In this proposal the ISL and Einstein's field equations are obtained from quantities such as energy, entropy, temperature and the counting of degrees of freedom which is set by the Holographic Principle \cite{tHooft,Susskind}. An implication of this formulation is that gravity is not a fundamental interaction, or at least, it is not directly related with the spin-2 state associated with the graviton, which is found in a fundamental theory such as string theory. 

We review here the basic features of the derivation of the ISL and refer the reader to Ref. \cite{Verlinde} to a discussion of the derivation of Einstein's field equations. 

Consider an holographic screen and a particle of mass, $m$, that travels at a distance $\Delta x$ in the external side of the screen where space has emerged. The entropy change associated to the information on the boundary, is assumed to be linear in the displacement $\Delta x$,
\be\label{Eq2.1}
\Delta S=2\pi k_B {\Delta x \over \lambda _c}~,
\ee
where $\lambda_c= \hbar/mc$ is the Compton wavelength and $k_B$ Boltzmann's constant. This result arises from saturation of the Bekenstein bound \cite{Bekenstein}, $\Delta S \le 2 \pi k_B E \Delta x / \hbar c$, $E$ being the relativistic energy. 

Consider now a mass $M$ and assume that its energy $U=Mc^2$ is projected onto a spherical holographic screen at a distance $r$ from $M$. The screen has area $A= 4 \pi r^2$ and it is divided in $N$ cells of the fundamental unit of area. This quantum of area $a_0$ is presumably associated with the square of Planck's length, $l_P$, that is, $a_0=l_P^2=G\hbar/c^3$. Thus, $N$ is given by:
\be\label{Eq2.1a}
N={A\over a_0}={{4\pi r^2} \over a_0}~.
\ee
This follows from the Holographic Principle, already mentioned above, according to which the number of degrees of freedom grows with the area and that the interior of a space can be described in terms of a theory on the boundary of the space.

The entropic force, that is, the force that arises from the Second Law of Thermodynamics, is evaluated as
\be\label{Eq2.2}
F \Delta x=T \Delta S~.
\ee

Considering that the surface is in thermal equilibrium at the temperature $T$, then all the $N$ bits have the same probability, and the energy of the surface is equipartitioned among them, as
\be\label{Eq2.5}
U={1\over2}Nk_B T~.
\ee
Therefore, substituting Eqs. (\ref{Eq2.1}), (\ref{Eq2.1a}), (\ref{Eq2.5}) and the expression for $U$ into Eq. (\ref{Eq2.2}), leads to Newton's law for gravity, $F=GMm/r^2$, a quite pleasing result.

As discussed in Ref. \cite{Verlinde}, Einstein's field equations follow from the assumption that a temperature can be associated with the gradient of a scalar quantity built out a Killing vector and the definition of the Komar mass. We refer the interested reader to that reference.

Let us now assume that holographic surface is described by a noncommutative geometry, more especifically, by a canonical phase-space noncommutative algebra. In two dimensions such a theory was studied in the context of the gravitational quantum well \cite{Bertolami1}, but one can easily generalize it to $d$ space dimensions:
\be\label{eq1.1}
\left[\hat q'_i, \hat q'_j \right] = i\theta_{ij} \hspace{0.2 cm}, \hspace{0.2 cm} \left[\hat q'_i, \hat p'_j \right] = i \hbar \delta_{ij} \hspace{0.2 cm},
\hspace{0.2 cm} \left[\hat p'_i, \hat p'_j \right] = i \eta_{ij} \hspace{0.2 cm}, \hspace{0.2 cm} i,j= 1, ... ,d
\ee
where $\eta_{ij}$ and $\theta_{ij}$ are antisymmetric real constant ($d \times d$) matrices and $\delta_{ij}$ is the identity matrix. This extended algebra is related to the standard Heisenberg-Weyl algebra
\be\label{eq1.2}
\left[\hat q_i, \hat q_j \right] = 0 \hspace{0.2 cm}, \hspace{0.2 cm} \left[\hat q_i, \hat p_j \right]
= i \hbar \delta_{ij} \hspace{0.2 cm}, \hspace{0.2 cm} \left[\hat p_i, \hat p_j \right] = 0 \hspace{0.2 cm},
\hspace{0.2 cm} i,j= 1, ... ,d ~,
\ee
by a class of linear non-canonical transformations between variables that, in physics, are usually called Seiberg-Witten map \cite{Seiberg-Witten}, while in the mathematics literature, the name Darboux map is preferred. These transformations are not unique. But it is important to point out that all physical predictions (expectation values, eigenvalues, probabilities) are independent of the choice of the map \cite{Bastos4}.

We expect that the underlying unit cell on the holographic screen is modified. We consider the following Ansatz \cite{Bastos11}: 
\be\label{Eq2.6}
a_0 \longrightarrow a_{NC}={G \hbar_{eff}\over c^3}~,
\ee
where $\hbar_{eff}$ is an effective Planck constant, which accounts for the noncommutative effects. Consequently, the temperature changes to
\be\label{Eq2.7}
T={G\hbar_{eff}\over ck_B}{M\over 2\pi r^2}~.
\ee
As the Compton wavelength does not correspond to an area, we assume that it remains unaltered and similarly
\be\label{Eq2.9}
\Delta S_{NC} = \Delta S= 2 \pi k_B {mc\over\hbar} \Delta x~.
\ee
Thus, the noncommutative correction to the force ($F_{NC}$) is such that
\be\label{Eq2.10}
{F_{NC} \over F}={\hbar_{eff} \over \hbar}~.
\ee

We still have to estimate the effective Planck constant. This can be regarded (up to a multiplicative constant) as the unit phase-space cell. For a two dimensional system the minimal phase-space cell has volume $\hbar^2$. In the present case, we must minimize the volume functional:
\be\label{Eq2.11}
V(\Delta x_1, \Delta p_1, \Delta x_2,\Delta p_2)=\Delta x_1 \Delta p_1 \Delta x_2 \Delta p_2~,
\ee
subject to the following constraints
\be\label{Eq2.12}
\begin{array}{l l}
\Delta x_1 \Delta p_1 \ge \hbar/2, & \Delta x_1 \Delta x_2 \ge \theta_{12}/2~,\\
& \\
\Delta x_2 \Delta p_2 \ge \hbar/2, & \Delta p_1 \Delta p_2 \ge \eta_{12}/2~.
\end{array}
\ee
This optimization problem should be solved by application of the Karush-Kuhn-Tucker Theorem \cite{Jahn}. However, this is actually an ill-posed problem since it is well-known that the quantum uncertainty relations (\ref{Eq2.12}) cannot be all saturated simultaneously. More precisely, a quantum state can saturate at most one of these conditions \cite{Bolonek,Kosinski}.

There is thus, no minimal volume of the phase-space cell. If we choose to minimize the product $\Delta x_1 \Delta x_2$, which corresponds to $\theta_{12}/2$, then the volume functional will be minimal for $\Delta p_1 \Delta p_2=\eta_{12}/2$. Altogether,
\be\label{Eq2.13}
V(\Delta x_1, \Delta p_1, \Delta x_2,\Delta p_2) \ge {\theta_{12}\eta_{12}\over4}~.
\ee
Similarly, if we choose to minimize the product $\Delta x_1 \Delta p_1$, corresponding to $\hbar/2$, then the volume functional will be minimal for $\Delta x_2 \Delta p_2 = \hbar /2$, and thus
\be\label{Eq2.14}
V(\Delta x_1, \Delta p_1, \Delta x_2,\Delta p_2) \ge {\hbar^2\over4}
\ee
An assumption to comply with Eqs. (\ref{Eq2.13}) and (\ref{Eq2.14}) is
\be\label{Eq2.15}
V(\Delta x_1, \Delta p_1, \Delta x_2,\Delta p_2) \ge {\hbar^2\over4} + {\theta_{12}\eta_{12}\over4} \equiv {\hbar_{eff}^2\over4}~,
\ee
with
\be\label{Eq2.16}
\hbar_{eff}= \hbar \left(1+ {\theta_{12}\eta_{12}\over\hbar^2} \right)^{{1\over2}}~.
\ee
Assuming that $\theta\eta / \hbar^2 << 1$, we finally obtain \cite{Bastos11}:
\be\label{Eq2.17}
F_{NC}= {GMm\over r^2} \left(1+ {\theta_{12}\eta_{12}\over2 \hbar^2} \right)~.
\ee
Thus, the force $F$ acquires a non-trivial correction if noncommutativity is not isotropic. This putative anisotropy implies that nocommutativity will affect the gravitational fall of masses on different directions. Assuming, for instance, that $\theta_{12}\eta_{12} \simeq O(1) \theta_{23}\eta_{23} \equiv \theta\eta$, then the Equivalence Principle, whose most stringent experimental limit for the differential acceleration of falling bodies, is expressed by $\Delta a/a \lsim10^{-13}$ \cite{Adelberger}, can be used to set a bound on the dimensionless quantity $\theta\eta / \hbar^2$:
\be
{\theta\eta\over \hbar^2} \lsim O(1) \times 10^{-13}~.
\label{Eq3.2}
\ee
This bound states that noncommutative effects can be 10 orders of magnitude greater than discussed in the context of the noncommutative gravitational quantum well \cite{Bertolami1} and the one arising from corrections to the hydrogen atom spectrum \cite{BertolamiQueiroz}. 

Another surprising implication of this result concerns a putative connection with the observed value of the cosmological constant. Indeed, it has been argued that a breaking of the Equivalence Principle at about $\Delta_{EP} \simeq10^{-14}$ level implies that the vacuum energy is related with the observed discrepancy between the observed value of the cosmological constant and the expected value from the electroweak symmetry breaking \cite{Bertolami10}:
\be
{\Lambda_{Obs.} \over \Lambda_{SM}} = {\Delta^4_{EP}}~,
\label{Eq4.1} 
\ee
and thus
\be
{\Lambda_{Obs.}\over \Lambda_{SM}} \simeq \left({\theta\eta\over \hbar^2}\right)^4 ~,
\label{Eq4.2}
\ee
a quite interesting relationship. Notice that the bound Eq. (\ref{Eq3.2}) is fairly close to the one suggested by Eq. (\ref{Eq4.1}).

\section{Emergence of our universe from the multiverse?}

The idea of a multiverse has been repeatedly discussed over the last fifty years. Everett's picture for the measurement problem \cite{Ev}, the MWI of QM, was most likely the first time that this radical concept was considered. The MWI states that upon a measurement by an observer, the wave function, splits into all the possible outputs, corresponding each one to an independent universe. In fact, this interpretation fits quite well with the process of decoherence: the interaction of the quantum system with its environment selects the base of the observables, such that the resulting density matrix describes a superposition of classical states. 

Later, in the early 1980's, a development on the inflationary model, lead to the idea of an eternally expanding de-Sitter space where the inflaton, the field responsible for the accelerated expansion of the early universe, sits in a metastable state and the ensued first order phase transition gives rise to the nucleation of bubbles of the "true" vacua, which can be seem as a universe with its own Hubble horizon \cite{Linde}. 

More recently, the multiverse reappeared in the context of string theory, when it was realized that the theory was actually a continuum of theories in the supermoduli-space. The number of possible vacua is huge, of order $G = 10^{100}$ or $10^G$! However, in our universe the observed value of cosmological constant is $10^{120}$ smaller than its natural value, $M_p^{4}$, $M_p$ being the Planck mass, therefore there must exist a huge number of universes with broken supersymmetry and all possible values for the cosmological constant \cite{BoussP, SussL, WeinM}. This scenario raises several intriguing questions: How is the vacuum of our world selected? Through anthropic arguments \cite{PolL}? Through quantum cosmology arguments \cite{Holman}? Is the string landscape scenario compatible with predictability \cite{Ellis}? Do the universes of the multiverse interact \cite{Bert2008}? Does the multiverse exhibit collective behavior \cite{Alonso}?

A quite interesting development concerns the possibility that these ideas about the multiverse are actually avatars of the same multiverse idea \cite{Bousso,Nom}. The key issue in this unification is to attach to an observer a local description of space-time. Thus, in order to describe the nucleation of bubble-universe, the suggestion is to follow the geodesic of an observer and to define, within its causal diamond, the decoherence process. This is performed such that the degrees of freedom that escape the future boundary of the causal diamond are inaccessible and are traced out. This leads to the branching of the wave function. Hence, the MWI is reconciled with an eternally inflating space-time.

However, this picture can only be consistent if the branches of the MWI do not interact or communicate with each other. Furthermore, even though, non-linear terms in the 
Schr\"odinger equation seem to be irrelevant at low energies \cite{Wein} (see also Ref. \cite{Bert2005}), non-linear terms are expected at higher energies from quantum cosmological \cite{Bertolami1991} or string field theory \cite{Banks} inspired arguments. In any case, it is likely that the mixture of low-energy and high-energy states seem to be a necessary ingredient to solve the cosmological constant problem, and thus this feature of high-energy models, might lead to non-linear extensions of quantum mechanics.The general structure of these non-linearities are of the type discussed in Refs. \cite{Wein}, which lead to EPR-like violations and communication between MWI branches \cite{Pol}. 

Therefore, on general grounds it is rather plausible that at the most fundamental level non-linearities are present in the evolution of the fundamental degrees of freedom describing the dynamics of an observer in space-time. If so, the EPR violations discussed in Ref. \cite{Pol} for non-linear observables are not only a challenge to the MWI, but they turn the unification proposed in Ref. \cite{Bousso} unviable too. This means that the idea that it suffices describing the dynamics of an observer to apprehend the physics of a single universe is untenable \cite{Bertolami-Herdeiro}.

\vspace{0.3cm}

\noindent
{\bf Acknowledgments} 

\vspace{0.3cm}

\noindent
This work of is partially supported by the Portuguese Funda\c c\~ao para Ci\^encia e a Tecnologia (FCT) under the project PTDC/FIS/11132/2009. The author would like to warmly thank the organizers for the kind invitation to talk at the Time Machine Factory and for the hospitality in Torino.

\end{document}